\documentclass[%
 reprint,
 amsmath,amssymb,
 aps,
]{revtex4-1}

  \usepackage[pdftex]{graphicx}
  \graphicspath{{../Figs/}}
  \DeclareGraphicsExtensions{.pdf,.png}

  \usepackage{bm}
  \usepackage{IEEEtrantools}
  \newcommand{\beq}{\begin{IEEEeqnarray}{rCl}}
  \newcommand{\eeq}{\end{IEEEeqnarray}}
  \DeclareMathOperator*{\argmax}{argmax}

  \usepackage[nomargin,inline,draft,author=]{fixme}
  \fxusetheme{color}
  \usepackage{siunitx}

  \newcommand{\threeast}{\bigskip\par\centerline{*\,*\,*}\medskip\par}%

\begin{document}
\preprint{APS/123-QED}

%%% PREAMBLE
  % \tableofcontents
  % \newpage
  % \listoffixmes
  % \newpage

  \title{A silicon photonic modulator neuron}
  % \title{Modulator-class photonic neurons based on a silicon microring}

  \author{Alexander~N.~Tait}
  \email{atait@ieee.org}
  \altaffiliation[Now at ]{National Institute of Standards and Technology, Boulder, CO 80305, USA.}
  \author{Thomas~Ferreira~de~Lima}
  \author{Mitchell~A.~Nahmias}
  \author{Heidi~B.~Miller}
  \altaffiliation[Also at ]{Department of Physics, Engineering Physics \& Astronomy, Queen's University, Kingston, ON K7L 3N6, Canada.}
  \author{Hsuan-Tung~Peng}
  \author{Bhavin~J.~Shastri}
  \altaffiliation[Also at ]{Department of Physics, Engineering Physics \& Astronomy, Queen's University, Kingston, ON K7L 3N6, Canada.}
  \author{Paul~R.~Prucnal}

  \affiliation{%
   Department of Electrical Engineering, Princeton University, Princeton, NJ 08544, USA
  }%
  \date{\today}

\begin{abstract}
  \noindent There has been a recently renewed interest in neuromorphic photonics, a field promising to access pivotal and unexplored regimes of machine intelligence. Progress has been made on isolated neurons and analog interconnects; nevertheless, this renewal has yet to produce a demonstration of a silicon photonic neuron capable of interacting with other like neurons. We report a modulator-class photonic neuron fabricated in a conventional silicon photonic process line. We demonstrate behaviors of transfer function configurability, fan-in, inhibition, time-resolved processing, and, crucially, autaptic cascadability -- a sufficient set of behaviors for a device to act as a neuron participating in a network of like neurons. The silicon photonic modulator neuron constitutes the final piece needed to make photonic neural networks fully integrated on currently available silicon photonic platforms.
\end{abstract}

\pacs{Valid PACS appear here}
% \ocis{(250.5300) Photonic integrated circuits; (070.1170) Analog optical signal processing; (130.7408) Wavelength filtering devices; (200.4700) Optical neural systems; (200.3050) Information processing.}

\maketitle

% \section{Motivation}
\section{Introduction}
  % General

    Renewed interest in neuromorphic photonics has been heralded by advances in photonic integration technology~\cite{Orcutt:12,Lim:14,Thomson:16}, roadblocks in conventional computing performance~\cite{Marr:13,Hasler2013}, the return of neuromorphic electronics~\cite{Merolla:2014,Furber:14,Benjamin:14,Meier:15,Davies:18}, and the inundation of machine learning (ML) with neural models~\cite{Schmidhuber:15}. Neural networks have held some role in ML (e.g. image and voice recognition, language translation, pattern detection, and others) since the 1950s~\cite{Neumann:56,rosenblatt58}. They fell out of favor in the 90's because they are difficult to train.

  % There is demand for neural networks
    Over the past decade, neural network models have decisively retaken the helm of ML under the alias of ``deep networks''~\cite{Najafabadi:15}. There are three main reasons: 1) major algorithmic innovations~\cite{LeCun:98,LeCun:15}, 2) the Internet: an inexhaustible source of millions of training examples, and 3) new hardware, specifically graphical processing units (GPUs)~\cite{Pallipuram:12}.
    Central processing units (CPUs) are woefully inefficient at evaluating these models because they are centralized and instruction-based, whereas networks are distributed and capable of adaptation without a programmer. GPUs are more parallel, but, today, even they have been pushed to their limits~\cite{Diamond:16}.

  % Neuromorphic electronics is back
    Today's demand for evaluating neural network models necessitates new hardware.
    High-tech juggernauts and research agencies have heavily invested in massively parallel application-specific integrated circuits (ASICs) for evaluating neural network models more efficiently, notably IBM~\cite{Merolla:2014}, HP~\cite{Pickett:13}, Intel~\cite{Davies:18}, Google~\cite{Jouppi:17,Graves:2016aa}, the Human Brain Project~\cite{Markram:12}, and DARPA SyNAPSE~\cite{Cassidy:13}. Some of these architectures aim to be ML number crunchers~\cite{Miyashita:17,Jouppi:17}, and others have enabled novel neuroscientific tools~\cite{Pfeil:13,Friedmann:13} and previously unforeseen low-power mobile applications~\cite{Tsai:16}.

  % Neuromorphic electronics are slow, but there are applications out there for fast ones
    The primary performance driver for the neuromorphic electronics community is computational power efficiency; speed is a secondary consideration. Neuromorphic electronics have largely focused on biological-timescale neural networks: kHz (with one 10MHz exception~\cite{Pfeil:13}). They universally rely on digital time- and event-multiplexing~\cite{Mundy:16}, which means they cannot simply run faster by turning up the clock. Nevertheless, there are compelling applications for neural networks with nanosecond latency. Some applications could be offline (i.e. number crunching) such as accelerators for deep network training and inference; others could be online (i.e. real-time) such as pattern detectors for wideband radio frequency (RF) signals and feedback controllers for systems subject to short-timeconstant instabilities.
    % I moved online and offline applications up here ^
    Moving beyond the nanosecond will require moving beyond purely electronic physics.

  % photonics can bring the fast, and some prior work
    Photonic physics exhibit properties distinct from those of electronics in terms of multiplexing, energy dissipation, and cross-talk. These properties are favorable for dense, high-bandwidth interconnects~\cite{Rakheja:2012} in addition to configurable analog signal processing~\cite{Weiner:11,Perez:17,Liu:16a}. Consequently, neuromorphic photonic systems could operate 6--8 orders-of-magnitude faster than neuromorphic electronics~\cite{Shastri:18} with potentially higher energy efficiencies~\cite{FerreiraDeLima:17}. Optical neural interconnects based on field evolution in free-space~\cite{Brunner:15,Psaltis:90}, holograms~\cite{Goodman:78,Asthana:93}, and fiber~\cite{Hill:2002} have been shown but were not widely adopted, in part because they cannot be integrated on a chip and thereby scaled robustly and manufactured cheaply. Analog interconnects integrated on a silicon photonic platform have been shown~\cite{Carolan:15,Shen:17}, but these interconnects are optically coherent, and a cascadable photonic neuron that regenerates phase from layer to layer has not been proposed.

    \threeast% adding an asterism before punchline

  % Punchline
    In this work, we fabricate and demonstrate a silicon photonic modulator neuron. It consists of a balanced photodetector directly connected to a microring (MRR) modulator. We demonstrate that this device possesses the necessary capabilities of a network-compatible neuron: fan-in, high-gain optical-to-optical nonlinearity, and indefinite cascadability -- properties never before demonstrated together in a single integrated device. Furthermore, we demonstrate optional, but useful, capabilities of transfer function configurability, inhibitory fan-in, pulse compression, and time-resolved processing. To establish empirical evidence of indefinite cascadability, we employ the methodology of~\cite{Tait:17}: inducing an observable bifurcation in an autapse circuit.

    Network-compatible photonic neurons are optical-in, optical-out devices that must be able to 1) convert multiple independently weighted inputs into a single output (i.e. \textit{fan-in}), 2) apply a nonlinear transfer function to the weighted sum of the inputs, and 3) produce an output capable of driving multiple other neurons, including itself (i.e. \textit{cascadability}). Research in neuromorphic photonics has seen an abundance of semiconductor lasers that exhibit nonlinear transfer functions, particularly those of a spiking neuron~\cite{Nahmias:13,coomans2011solitary,VanVaerenbergh:13,Brunstein:2012,Selmi:2014,Romeira:16,Nahmias:16,Deng:17,Peng:18}, reviewed in~\cite{Prucnal:16advances}. Nevertheless, all of the conditions of network-compatibility have yet to have been conclusively demonstrated in a single device (with exception of the fiber laser in~\cite{Shastri:2016aa}), and much of this research overlooks fan-in and/or cascadability entirely. There are fundamental reasons why fan-in, nonlinearity, and cascadability are difficult to achieve in all-optoelectronic devices, discussed further in Sec.~\ref{sec:claiming-cascadability}. The modulator neuron addresses these issues using an optical/electrical/optical (O/E/O) signal pathway with a lumped, intermediate electronic connection.

\begin{figure}[tb]
      \begin{center}
      \includegraphics[width=0.99\linewidth]{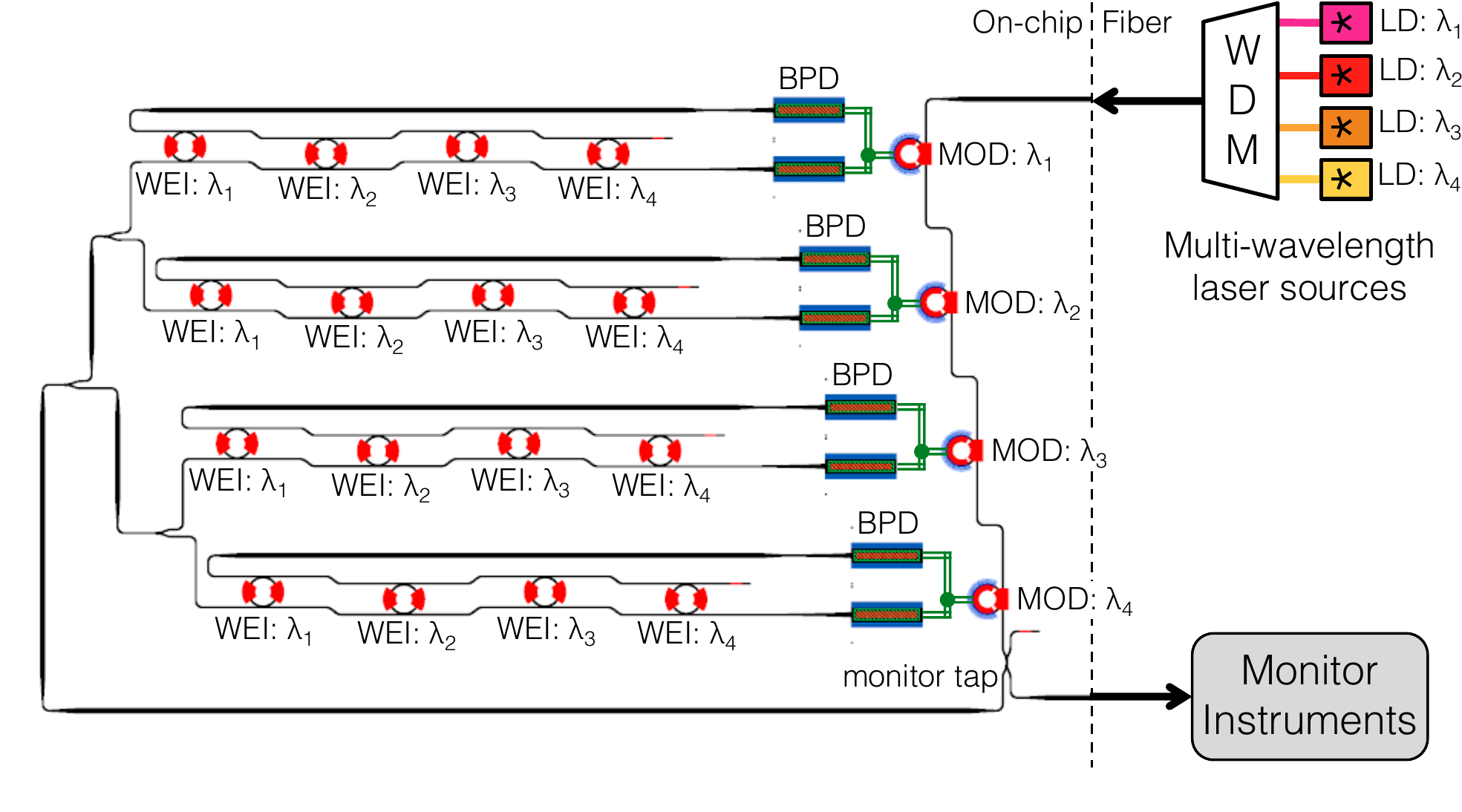}
      \caption[]{A broadcast-and-weight network~\cite{Tait:14} using MRR modulator neurons and MRR weight banks~\cite{Tait:18fb}. The entire network is integrated with the exception of pump lasers, which lie outside of the high-bandwidth signal pathway.}
      \label{fig:conceptNetwork}
      \end{center}
    \end{figure}

  % compatibility with weight banks
    The neuron shown here is integrable with the silicon photonic neural network demonstrated in~\cite{Tait:17}. Modulator neurons and the MRR weight network combine to form a neuromorphic photonic architecture called broadcast-and-weight~\cite{Prucnal:17,Tait:14}, diagrammed in Fig.~\ref{fig:conceptNetwork}. In this architecture, photonic neurons output optical signals with unique wavelengths. These are wavelength-division multiplexed (WDM) and broadcast to all others, weighted, and detected. Each connection between a pair of neurons is configured independently by one MRR weight~\cite{Tait:16scale,Tait:16multi}, and the WDM carriers do not mutually interfere when detected by a single photodetector. In the context of prior work, the MRR modulator neuron illustrates a path to fully integrated photonic neural networks compatible with currently available silicon technology. This compatibility is a crucial element of scalability, cost, and feasibility.

  % Silicon photonics offers a solution
    Silicon photonics presents an opportunity for optical technologies to venture beyond pure communication links and specialized co-processing boxes. Recent advances have brought about a stable academic/industrial ecosystem surrounding silicon photonics~\cite{Soref:87,Soref:10}, propelling technology road-mapping~\cite{Thomson:16}, standardized multi-project wafer runs~\cite{Lim:14,Orcutt:12}, broadened accessibility to academic research~\cite{Chrostowski:15}, and economies of scale previously enjoyed solely by microelectronics. Despite its origins in cutting the cost of mid-reach communications, silicon photonic integration could be instrumental to large-scale photonic information processing concepts -- beyond what could be considered in fiber, III-V, or holographic platforms.

\section{Methods}
  \subsection{Device Description}

\begin{figure}[tb]
      \begin{center}
      \includegraphics[width=0.99\linewidth]{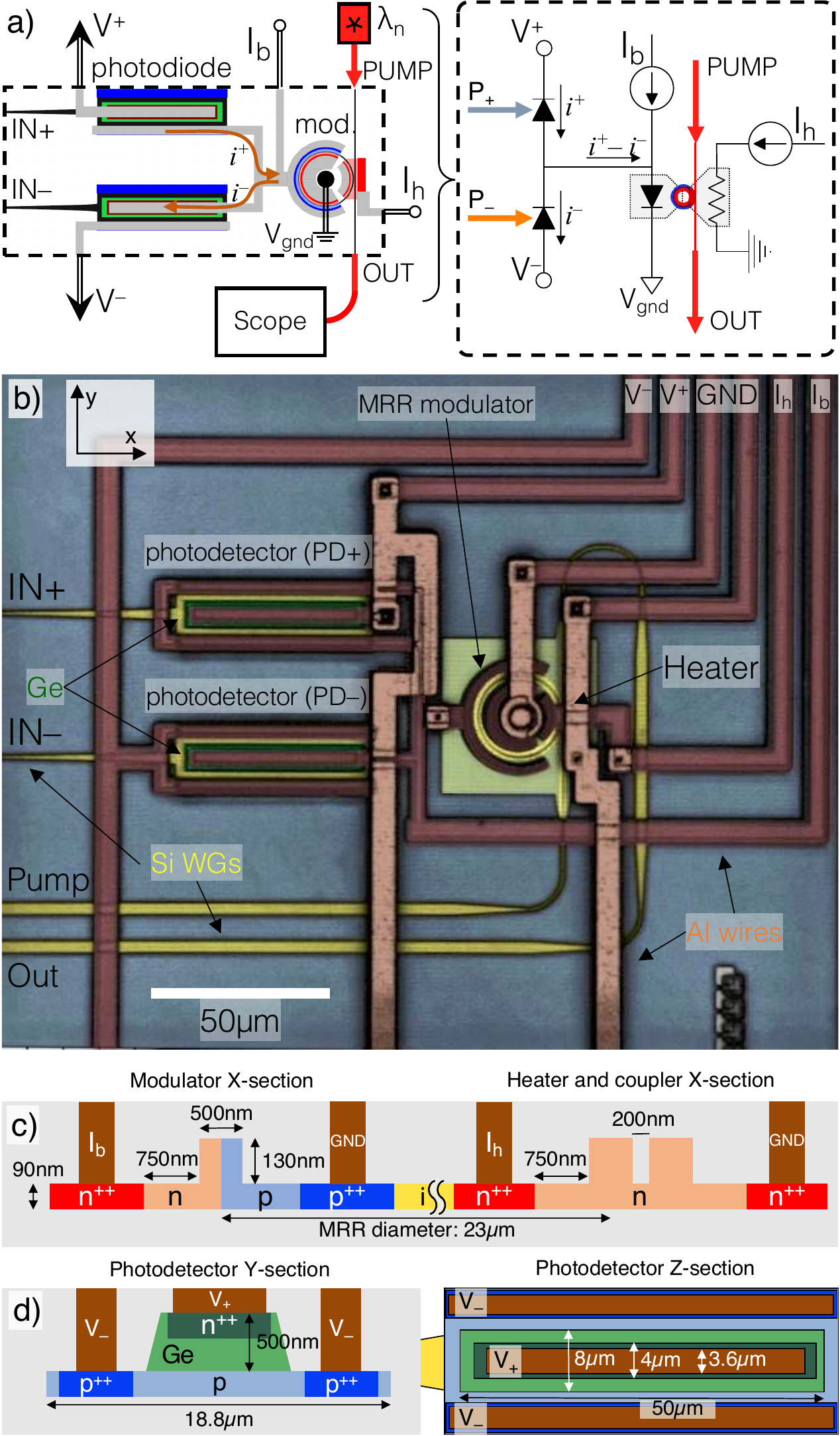}
      \caption[]{a) Simplified equivalent circuit diagram of the MRR modulator neuron. b) False color confocal micrograph of the fabricated device. c) Cross-section of the MRR modulator with embedded PN modulator and N-doped heater. d) Cross-section of Si-Ge photodetector, showing cathode ($V_+$) and anode($V_-$).}
      \label{fig:device-image}
      \end{center}
    \end{figure}

    \paragraph{Theory of operation}
    The modulator neuron is an optical-to-electrical-to-optical (O/E/O) device consisting of two balanced photodetectors (PDs) connected electrically to a microring (MRR) modulator. It that takes two balanced optical inputs, subtracts them electronically, and remodulates a signal onto a new wavelength~(Fig.~\ref{fig:device-image}(a)). The output signal is a nonlinear function of inputs determined by the electro-optic transfer function of the modulator.

    Fig.~\ref{fig:device-image}(a) shows the neuron circuit diagram. The WG labeled ``pump'' directs a continuous-wave (CW) laser signal at $\lambda_n$=1544.8nm to the MRR modulator neuron. When modulated, this signal serves as the neuron's optical output. The PDs are reverse-biased and convert impinging light at all wavelengths into photocurrents. Photocurrent from the positive PD adds to the injected current, while photocurrent on the negative PD shunts away some of the injected current coming from the bias port. These photocurrents are combined with a bias current $I_b$ such that their sum affects the refractive index of the MRR modulator via free-carrier injection. This index change affects its transmission at the pump wavelength, and ultimately modulates the amount of light transmitted at the output port of the neuron. The MRR resonance wavelength can also be thermally tuned by a in-situ heater with a current $I_h$, used for coarse wavelength locking and fine bias control.

    \paragraph{Fabrication}
    Fig.~\ref{fig:device-image}(b) shows a false-color optical micrograph of the fabricated neuron. Waveguides (WGs), shown in yellow, are formed in the 220nm thick silicon layer. They are patterned to be 500nm wide except in long, straight segments, WGs are widened to 3$\mu$m to minimize sidewall scattering. The MRR modulator consists of a circular WG with a designed radius of 11.5$\mu$m. It is coupled to one WG with a gap of 200nm. The WGs containing active devices are partially etched to a 90nm thick pedestal (light green), which hosts the dopants.

    Doping profiles of active regions are diagrammed in Fig.~\ref{fig:device-image}(c). The MRR has two doped regions: an N$^{+}$/N/N$^{+}$ junction to act as an ohmic heater and an N$^{+}$/N/P/P$^{+}$ junction to act as a high-speed modulator. Phosphorous and boron concentrations are the same as in~\cite{BaehrJones:12}.
    % were N: 5$\times 10^{17}$ and N+: 5$\times 10^{20}$ cm$^{-3}$. Boron dopant concentrations were P: 7$\times 10^{17}$ and P+: 1.7$\times 10^{20}$~\cite{BaehrJones:12}.
    A germanium layer is deposited and patterned (dark green) along with additional dopants that create a vertical P-I-N photodiode (Fig.~\ref{fig:device-image}(d)). Silicon oxide is then deposited, vias are etched, and aluminum traces are deposited (red) -- this sequence is repeated to form a second wiring layer (orange). Metal traces are patterned to break out the device electrical ports to a row of probe pads.

  \subsection{Device characterization} \label{sec:characterization}
    \begin{figure}[tb]
      \begin{center}
      \includegraphics[width=0.99\linewidth]{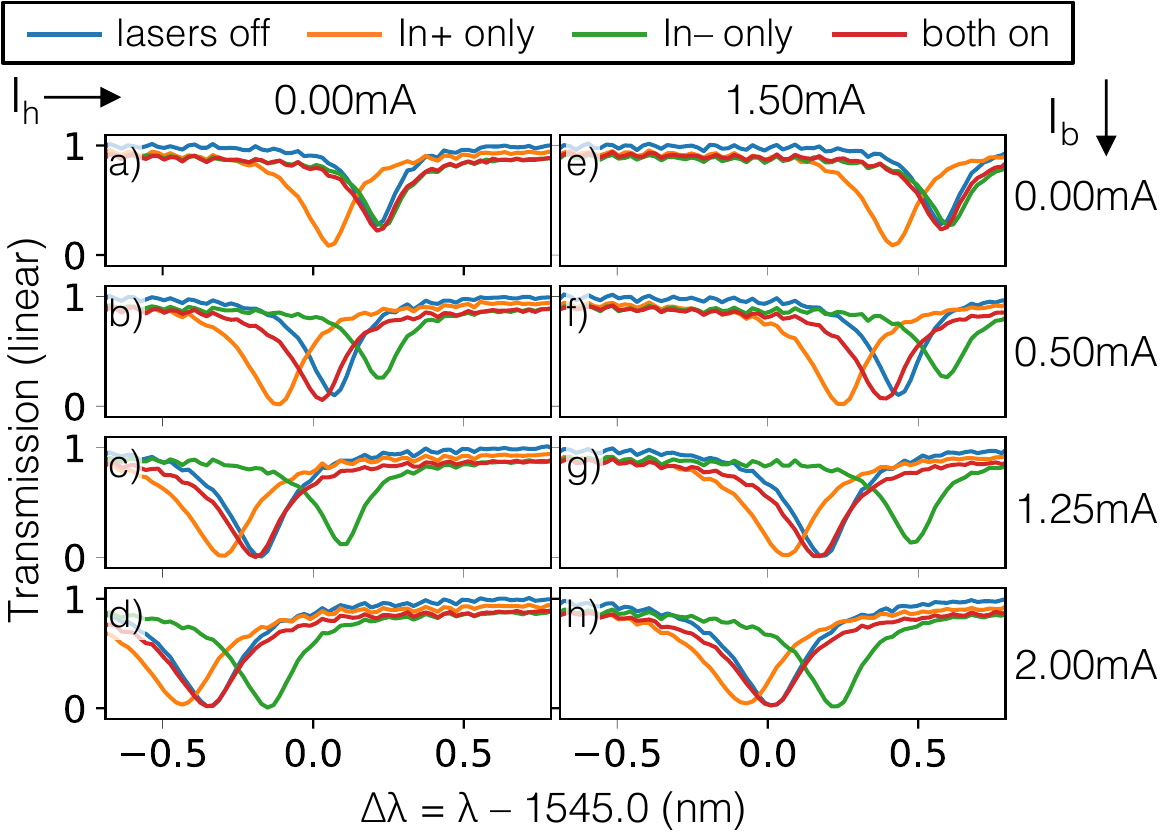}
      \caption[]{Spectrum vs. heat and forward current bias under different permutations of illumination. Columns distinguish heating; rows distinguish modulator bias current. Traces distinguish optical input conditions. Blue: no light, just electrical bias. Orange, green: only one optical input at a time. Red: Both inputs creating complementary effects. Ideally, the red traces should overlap with the blue.
      }
      \label{fig:spectra}
      \end{center}
    \end{figure}

    The optical behavior of the MRR modulator and its response to different effects is characterized by the profile of its resonance dip. We examine the optical transmission spectrum of the MRR modulator in response to three independent quantities:
    \begin{enumerate}
      \item Heater current bias ($I_h$)
      \item Modulator current bias ($I_b$)
      \item Optical power into the IN+ and IN-- ports ($P_+$ and $P_-$, respectively)
      % \fxnote{There's a specific term we could borrow from optical links. OMA: optical modulation amplitude. Thoughts?}
    \end{enumerate}
    The spectra are plotted in Fig.~\ref{fig:spectra}, taken by a transmission spectrum analyzer (Apex AP2440A). The bare resonance is seen in Fig.~\ref{fig:spectra}(a) as the blue curve. The modulator's PN junction is forward-biased by currents $I_b$ such that dynamic inputs induce an index change through free carrier injection. When electrical bias is applied, the peak blue shifts to the left, seen over different rows. When heat is applied, it red shifts to the right, seen over the different columns. Orange, green, and red curves show different optical input states, described below.

    Comparing the blue curves of (a-d) with (e-h), we see that thermal tuning does not significantly affect the response besides shifting the absolute wavelength with efficiency of 0.24nm/mW. This is desirable so that the thermal degree of freedom can be used as an independent parameter to lock the MRR onto a WDM channel of interest without affecting the electro-optic response. In contrast, the electrical bias parameter does effect the depth and quality factor (from 14.5k to 3.5k) of the peak. As a result, the electrical bias can be used to configure the response, while the combination of electrical and thermal parameters can be used to maintain the desired wavelength of interest. The electrical tuning efficiency below diode threshold is zero because no free carriers are injected. Above diode threshold, the tuning efficiency is at most 0.26nm/mA.

    \begin{figure*}[htb]
      \begin{center}
      \includegraphics[width=.72\linewidth]{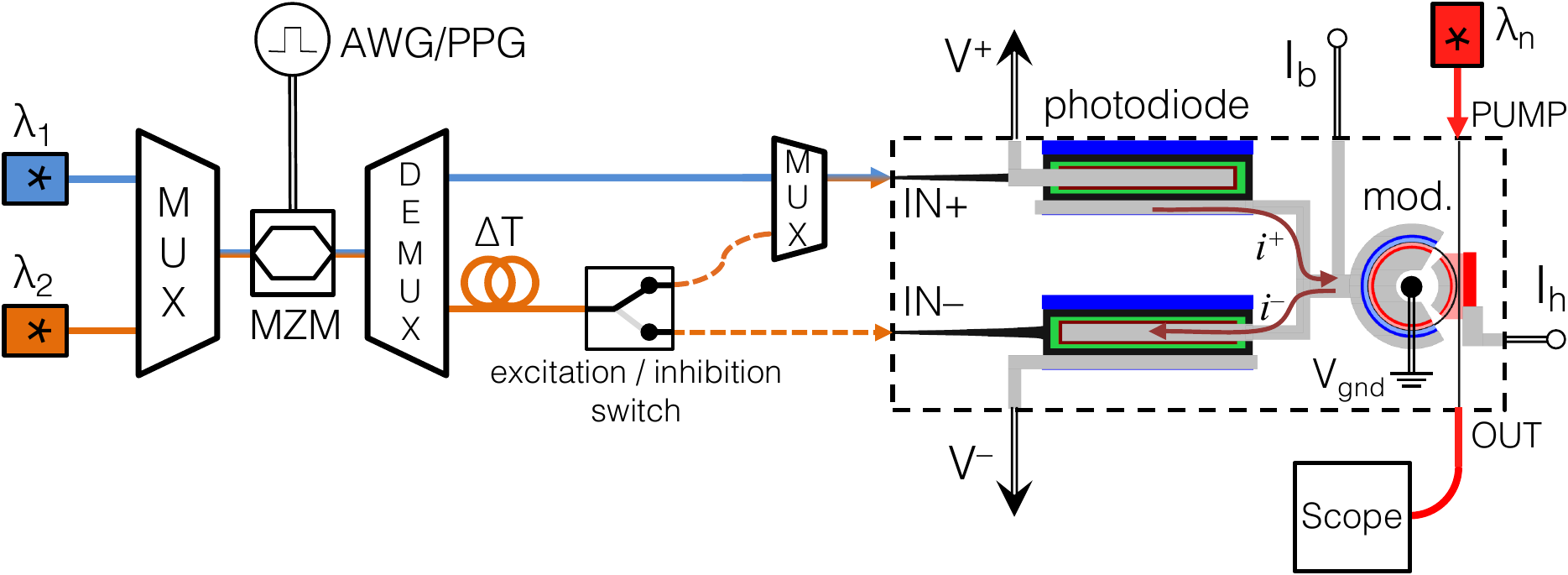}
      \caption[]{Experimental setup. A Mach-Zehnder modulator (MZM) followed by a DEMUX and $\Delta T$ delay is used to create two distinct signals. An arbitrary waveform generator (AWG) or a pulse-pattern generator (PPG) is used to drive the modulator. A fiber switch directs the second ($\lambda_2$) input either to the same IN+ port as the first input ($\lambda_1$) or to the complementary IN-- port. Light going into IN(+/--) waveguides with power $P_{+/-}$ impinge on photodetectors PD(+/--), resulting in complementary photocurrents $i^{(+/-)}$. Current injected to the MRR modulator (mod.) is the sum of photocurrents and the bias current ($I_b$): $I_b + i^+ - i^-$. MRR resonance wavelength is also affected by a heater biased at $I_h$. The transmission of the modulator is probed by a continuous-wave PUMP at $\lambda_n$. Its optical output is monitored by an oscilloscope (Scope). The dashed box indicates the on-chip/off-chip boundary.
      }
      \label{fig:the-setup}
      \end{center}
    \end{figure*}

    Two lasers are used to characterize the O/E/O response. When optical inputs are non-zero, the effect is the same as modifying the electrical bias because they deflect the current seen on the modulator. Orange curves represent optical power (10.dBm at facet) on the IN+ port with the IN-- laser turned off. The photocurrent adds to the bias current, causing a blue-shift. The fiber-to-fiber insertion loss through the chip is 18.dB, so we estimate that the input GC represents 9.0dB of loss. Therefore, the optical power reaching PD+ would be 1.26mW. This causes a deflection equivalent to 0.96mA, resulting in an estimated PD responsivity of 0.76A/W, slightly less than in other reports~\cite{Fard:16}. The green curve is the opposite case: IN+ laser off and IN-- laser at 10.dBm at facet. The red curve is with both lasers on. The fact that red curves match up with blue (lasers off) curves is significant because it shows that the +/-- optical inputs cancel out. This means that excitatory and inhibitory effects can be realized by switching the port of impinging inputs. Microring weight banks~\cite{Tait:16scale,Tait:18fb} explored in other work were specifically designed to perform this task of directing multiple wavelength channels between the ports of a balanced photodetector.

    The optical-to-optical gain is measured by comparing the amplitude of the modulation swing on the $\lambda_1$ input to the amplitude of the modulation swing on the $\lambda_n$ output coming from the neuron.
    It is found to be 2.16$\times 10^{-2}$, dominated by fiber-to-chip insertion loss measured at 18.0dB. Removing the coupling loss results in an on-chip gain of 1.36. Gain scales in proportion to the pump power, which was held artificially low during this measurement to eliminate the chance of the pump influencing the carrier concentration or temperature of the modulator. Pump power was measured as --5.5dBm at the chip facet, an estimated --14.5dBm arriving at the neuron, which is less than one hundred times weaker than it could be under realistic assumptions.

    We use injection modulation because we found that the depletion effect was too weak to demonstrate the desired results with the non-optimized modulators we fabricated. Injection modulation with a forward-biased junction is slower than depletion modulation with a reverse-biased junction. Injection modulation bandwidth can reach up to 6.25GHz~\cite{Xu:07} compared to depletion modulators shown up to 40GHz~\cite{Novack:13}. All of the signal processing concepts of modulator-class neurons shown here hold over different types of modulation mechanisms.

    % Not showing frequency response anymore
    % Frequency response of the photodetector-modulator system is assessed by applying a sinusoid modulation to an optical input. A laser pump is inserted to the modulator, and the modulator is electrically biased to have a deep dip and thermally biased such that the pump wavelength hits the maximum slope point. Frequency response is not assessed through direct electrical modulation because the frequency response of the long biasing wire would confound the result. In contrast, the external Mach-Zehnder modulation of the optical input has a bandwidth in the 10's of gigahertz. \fxwarning{This is all shown in Fig.???}

  \subsection{Experimental setup}
    The multi-channel input generator used in the below experiments is shown in Fig.~\ref{fig:the-setup}. All time-varying signals coming in and out of the chip are optical. Two input wavelengths ($\lambda_1$=1546.4nm and $\lambda_2$=1548.2nm) are wavelength-division multiplexed and power modulated by a fiber Mach-Zehnder modulator (MZM). Distinct input signals are then created by a relative delay, $\Delta T$, between these wavelengths following the method of~\cite{Tait:15}. Laser pumps for the external inputs and the neuron itself come from a distributed feedback (DFB) laser array (ILX 7900B). The $\lambda_2$ signal can be switched between the excitatory and inhibitory ports.

    There are three signal generators used in the below experiments, two analog (a.k.a. synths) and one binary. A simple, slow waveform generator (HP 8116A) is used to acquire transfer functions (Sec.~\ref{sec:transfer-functions}) and autapse behavior (Sec.~\ref{sec:cascadability-demo}). The 8116A offers control of sawtooth waveforms that can be used to separate rising and falling aspects. Burst inputs are generated by a Rohde and Schwartz SMBV 100A VG (R\&S), which is used in Secs.~\ref{sec:transfer-functions}, \ref{sec:continuous-time}, and \ref{sec:result-fanin}. The R\&S effectively yields trains of return-to-zero (RZ) pulses of varying amplitude. Binary pulsed inputs used in Sec.~\ref{sec:result-pulses} are generated by a pulse pattern generator (PPG) (Anritsu MP1761B). The PPG provides the highest instantaneous bandwidth but the least control over waveforms.

    The neuron's output is coupled off-chip, detected, and observed in a sampling oscilloscope (Tektronix DSA8300). Between the output coupler and scope, there is a signal-to-noise enhancement stage, not diagrammed, consisting of an erbium doped fiber amplifier (EDFA), optical bandpass filter at $\lambda_n$, discrete photodetector (Discovery Semiconductors, Inc. DSC-R405ER), and, for low-bandwidth experiments, an electrical low-pass filter. The pre-chip subsystem contains two EDFAs and polarization controllers that are not shown. All of the above instrumentation is controlled remotely via \textit{lightlab}, a free software python package~\cite{lightlab:105}.

    The chip containing the neuron is placed on a temperature-controlled alignment stage and kept in place by a back-side vacuum. 4 fibers of a V-groove array are aligned to grating couplers leading to the IN+, IN--, PUMP, and OUT ports. DC bias signals are applied through 6 probe tips in an array (GGB MCW-26-8146). Power and ground signals are derived from a power supply, and tunable biases, $I_b$ and $I_h$, are derived from two Keithley 2400 current sources.

\section{Results}
  We conduct a series of experiments on this silicon photonic device to support the claim that it exhibits the three properties necessary to act as a network-compatible neuron (nonlinearity, fan-in, cascadability), in addition to several optional but useful properties. Sec.~\ref{sec:transfer-functions} demonstrates nonlinear optical-to-optical conversion and transfer function configurability. Sec.~\ref{sec:continuous-time} demonstrates high-bandwidth operation and pulse compression. Sec.~\ref{sec:result-fanin} demonstrates fan-in and inhibitory fan-in. Sec.~\ref{sec:result-pulses} demonstrates time-resolved spike processing that is excitatory and inhibitory. Sec.~\ref{sec:cascadability-demo} demonstrates indefinite cascadability through the observation of bistability in an autapse circuit.

  \subsection{Transfer functions} \label{sec:transfer-functions}
    The photodetector-modulator system exhibits a variety of nonlinear optical-to-optical transfer functions that are relevant for a wide variety neural processing tasks. To obtain these responses, the optical input is modulated by a slow rising sawtooth waveform at 200kHz, derived from the HP synth. The type of response depends on the wavelength offset between the pump and the MRR resonance. To change the response, the MRR resonance is shifted by varying the heater bias. Corresponding time-domain behavior is then shown using a 100MHz burst from the R\&S generator.

    \begin{figure*}[tb]
      \begin{center}
      \includegraphics[width=0.8\linewidth]{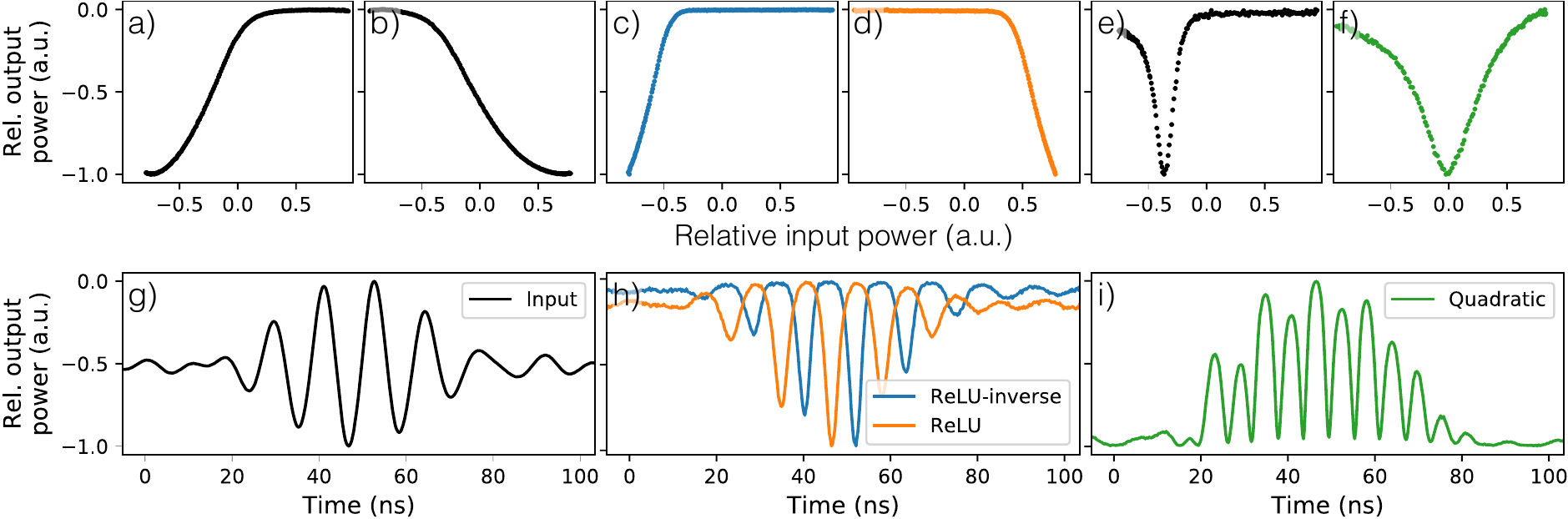}
      \caption[]{A variety of relevant O/E/O transfer functions seen from the PD-modulator pair taken at different heater bias conditions. a,b) Sigmoid, c,d) Rectified linear unit (ReLU), e) Radial basis function (RBF), and f) Quadratic. Below, time resolved pictures of these transfer functions: g) input or linear, h) both ReLUs, i) Quadratic. The input is a 40ns burst of a 100MHz carrier. All plots display experimental data.}
      \label{fig:transferFunctions}
      \end{center}
    \end{figure*}

    \begin{figure*}[htb]
      \begin{center}
      \includegraphics[width=0.9\linewidth]{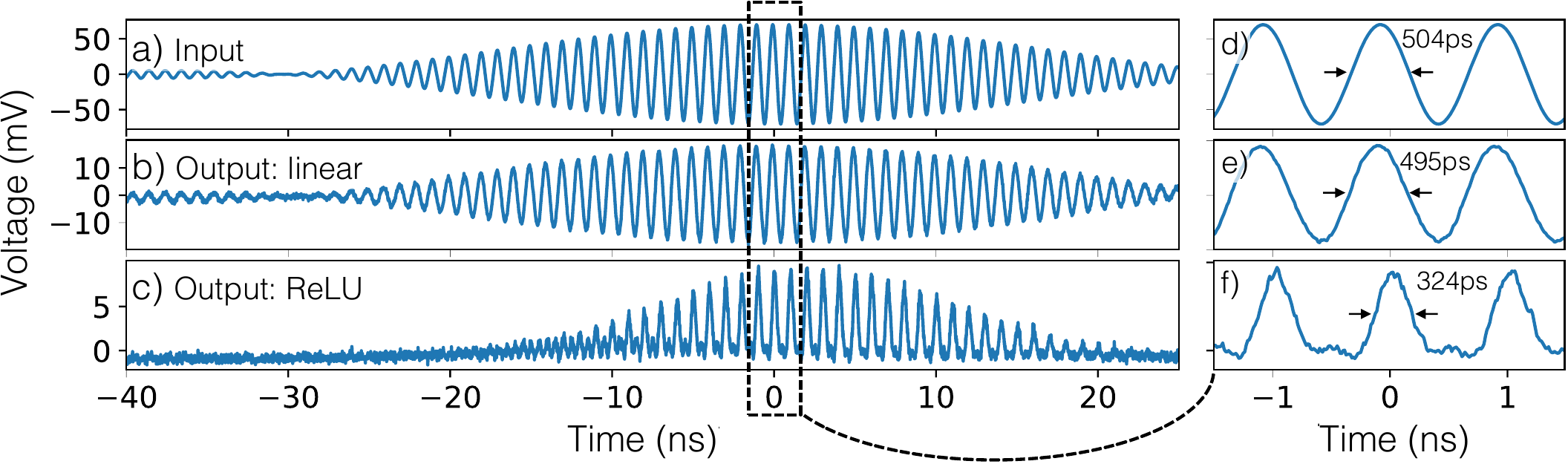}
      \caption[]{Reproduction, rectification, and pulse compression. a) The input is a 25.0ns burst of a 1.0GHz RF carrier modulated on a $\lambda_1$ optical carrier. b) In a linear regime, the modulating signal is faithfully reproduced on a different carrier wavelength, $\lambda_n$.}
      \label{fig:1GHzRectify}
      \end{center}
    \end{figure*}

    Fig.~\ref{fig:transferFunctions} shows six response shapes, each relevant in different areas of neural processing and machine learning. They are obtained under different biasing conditions.
    This variety indicates that the MRR modulator neuron can yield nonlinear configurability, to complement the configurability of the network's linear weights. The sigmoid shapes of Figs.~\ref{fig:transferFunctions}(a,~b) are commonly used in recurrent Hopfield networks for nonlinear optimization~\cite{Hopfield:85}. They are obtained by biasing at the maximum slope point. The rectified linear unit (ReLU) shape of Fig.~\ref{fig:transferFunctions}(c,~d) is widely used in feedforward machine learning networks today, i.e. in multi-layer perceptrons (MLPs) and convolutional neural networks (CNNs)~\cite{Jouppi:17}. Positive and negative ReLUs are obtained by biasing slightly off-resonance, either above or below the pump wavelength. A network that combines sigmoid and ReLU neurons is well-suited to solve nonlinear optimization problems with constraints, some of which are reviewed in~\cite{Wen:09}. The peaked transfer function of Figs.~\ref{fig:transferFunctions}(e,~f) are known as radial basis functions (RBFs). When biased on-resonance, the RBF is centered at zero, resulting in a quadratic or rectifying transfer function. The off-centered RBF is obtained by biasing off-resonance with high quality factor and a strong input amplitude. RBFs are commonly used for ML based on support vector machines~\cite{Cortes:95}.

    These nonlinear responses stem from the electro-optic transfer function of the modulator. The photodetectors' O/E response is linear and the modulator's E/O response is a strongly nonlinear, peaked Lorentzian.
    % Using the intermediate electrical junction contrasts with approaches that use direct, weak optical nonlinearities~\cite{Hill:2002} or resonator techniques that enhance optical field intensity at the expense of wavelength constraints that preclude WDM fan-in~\cite{Selmi:2014,VanVaerenbergh:13}. Since the electronic link is short and not used for inter-neuron communication, it does not degrade bandwidth~\cite{Nahmias:16}.
    It is apparent that the different responses correspond to different pieces of the MRR modulator's Lorentzian peak shape. To some extent, the type of response observed depends on the amplitude of the input, which is not ideal. The ideal ReLU continues increasing indefinitely, and the ideal sigmoid saturates indefinitely. We note however, that every neuron has some limitation to its range of inputs, whether it be imposed by register bit depth (digital) or state bleaching (analog, including biological).
    % Furthermore, the modulator neuron cannot exhibit the type of discontinuous response characteristic of Type II spiking neurons. Excitable lasers can exhibit such a response, as shown in Figs. 4 and 5 of~\cite{Shastri:2016aa}.
    % \fxwarning{this last paragraph lost me. Type II spiking neurosn were not previously mentioned and I am not sure what you want to say. I would just focus on the fact that it is still possible to train networks with global constraints such as max. output power. I have a reference for that in the other paper.}

  \subsection{Response to high-bandwidth inputs} \label{sec:continuous-time}
    % Fig.~\ref{fig:100MHzBurst} shows a time-resolved measurement of the MRR photonic neuron. It was taken with the single-input setup of Fig.~\ref{fig:all-setups}(a) with a 40.0ns burst of a 100MHz carrier, derived from the Rohde and Schwartz generator. The nonlinear responses of Figs.~\ref{fig:100MHzBurst}(b, c, d) are clearly correspondent with the transfer functions of, respectively, Figs.~\ref{fig:transferFunctions}(a, e, d). The linear response of Fig.~\ref{fig:100MHzBurst}(a) is obtained by biasing on the maximum slope point and driving with a signal weaker than that needed to resolve the sigmoid of Fig.~\ref{fig:transferFunctions}(d).
    % \begin{figure}[tb]
    %   \begin{center}
    %   \includegraphics[width=0.99\linewidth]{100MHzBurst}
    %   \caption[]{Time domain view of tunable transfer function. \fxwarning{the last one is too strong}}
    %   \label{fig:100MHzBurst}
    %   \end{center}
    % \end{figure}

    The response of the neuron to a much faster input is shown in Fig.~\ref{fig:1GHzRectify}. The input RF envelope is a 25.0ns pulse of a 1.0GHz carrier, which is modulated on an optical carrier at $\lambda_1$. The outputs indicate that, under different biasing conditions, the neuron can either reproduce or apply nonlinear transformations to a fast input. The linear response of Fig.~\ref{fig:1GHzRectify}(b) is obtained by biasing on the maximum slope point and driving with a signal weaker than that needed to saturate the sigmoid of Fig.~\ref{fig:transferFunctions}(b). This demonstration is significant, in part, because it is a faithful conversion of an RF signal from one wavelength, $\lambda_1$, to another, $\lambda_n$. A 20\% voltage gain is shown, although this is not necessarily representative of the fully integrated case because gain is affected by fiber-to-chip insertion loss and fiber EDFA configuration.

    In Fig.~\ref{fig:1GHzRectify}(c), the ReLU of Fig.~\ref{fig:transferFunctions}(b) was biased below its elbow. Viewed on the burst level, it exhibits a thresholding phenomenon resulting in a compressed burst width of 20.4ns (19\% compression). As seen in the zoomed in traces of Fig.~\ref{fig:1GHzRectify}(d-f), the ReLU also exhibits a pulse compression effect. The 1GHz input can be viewed as a train of 500ps full-width half maximum (FWHM) pulses that are turned into a train of 324ps pulses (36\% compression).

    Pulse compression is an important behavior for maintaining the integrity of pulses cascaded over multiple stages of neurons. It occurs because the leading and lagging pulse edges take on values that lie below the ReLU elbow. Both of these thresholding/compression behaviors are due to the positive third-order nonlinearity possessed by the ReLU. It is more common for optical devices to exhibit saturating behaviors with negative third-order nonlinearities, for example, those subject to two-photon absorption and laser diodes well above the lasing threshold. If saturating devices were used in a neural network, the pulse FWHM from stage-to-stage would increase (i.e. degrade) indefinitely.

    % The radio frequency (RF) spectra of the linear and ReLU outputs are shown in Fig.~\ref{fig:1GHzRectify}(g). \fxwarning{no it isn't. should it be added in? its a pretty small panel} These spectra are obtained by applying a Fourier transform to the time traces. The linear response shows that the neuron can faithfully reproduce the input at this frequency with a small cubic saturation visible at 3.0GHz in the frequency response. The ReLU response exhibits approximately half as much power at the fundamental frequency and more power at the second harmonic. This is, as expected, indicative of a nonlinear rectification behavior.
    % Leave it as backup for reviewer's concerns or supplementary material. --TFL

  \subsection{Response to multiple inputs} \label{sec:result-fanin}
    We then consider the case of two inputs at different wavelengths, $\lambda_1$ and $\lambda_2$. We demonstrate fan-in behavior across wavelengths -- the ability to perform a summation operation across multiple wavelengths while simultaneously converting them to a single wavelength, $\lambda_n$. Fan-in is central to the idea of network-based processing, so this feature is particularly important to demonstrate directly in order to claim a device as a ``photonic neuron.''

    \begin{figure}[tb]
      \begin{center}
      \includegraphics[width=0.99\linewidth]{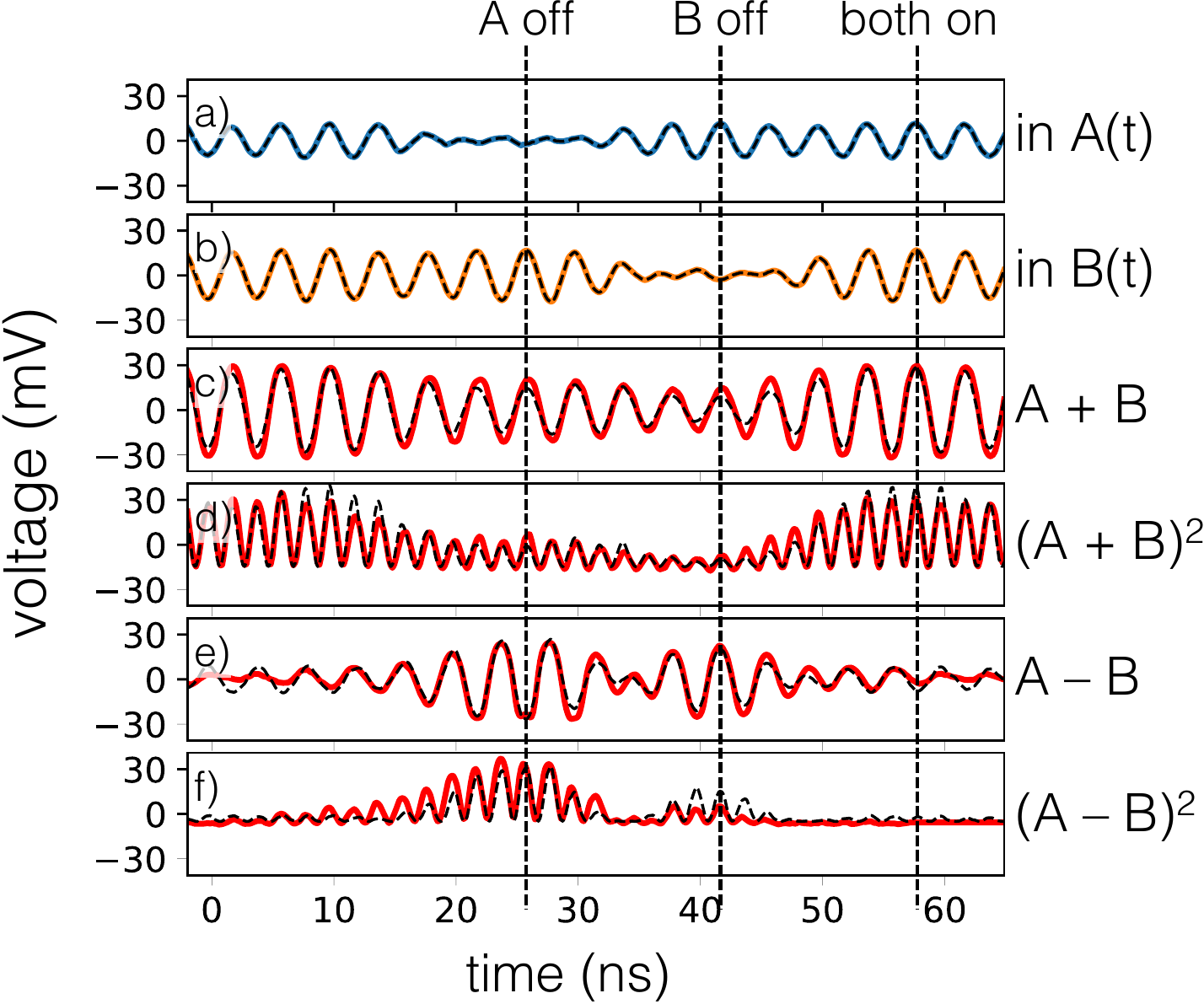}
      \caption[]{Burst addition and two channel rectification. Solid lines are experimentally measured; their colors correspond to the wavelengths in Fig.~\ref{fig:the-setup}. a,b) Inputs. c-f) Outputs under different biasing configurations performing the two-variable functions on the right. Dashed black curves are ideal outputs that are calculated from the measured inputs. The functions shown demonstrate linear and nonlinear, excitatory and inhibitory capabilities. Input~$B$ is slightly stronger than input~$A$.}
      \label{fig:two-burst}
      \end{center}
    \end{figure}

    The setup for addition and subtraction are shown in Fig.~\ref{fig:the-setup}. The input envelopes are delayed versions of one another, i.e. $B(t) = A(t + \Delta T)$. For subtraction, the two inputs are sent into complementary ports of the neuron's balanced PD. For addition, they are multiplexed and sent into the same port. This is represented by the excitatory/inhibitory switch in Fig.~\ref{fig:the-setup}. The modulating signal, a 900MHz RF carrier that is amplitude modulated (AM) at 50MBaud, is chosen to demonstrate several behaviors of interest to RF signal processing, discussed below. The delay, $\Delta T$, is adjusted so that the RF carrier waves are in-phase and the modulating bit pattern of [1, 1, 0] is delayed by one bit period.

    Figure~\ref{fig:two-burst}(a,b) shows the inputs $A(t)$ and $B(t)$ after detection. Figure~\ref{fig:two-burst}(c) is the optical output of the MRR neuron when these signals are multiplexed into the IN+ port. This illustrates basic optical fan-in of two WDM signals onto a single wavelength. Figure~\ref{fig:two-burst}(d) is a more complex case showing excitatory fan-in followed by a nonlinear rectifying conversion from the electronic to optical domain. Figure~\ref{fig:two-burst}(e, f) illustrates inhibitory behavior where the inputs counteract one another -- more than just an inversion of the excitatory case. Figure~\ref{fig:two-burst}(e) is with a linear neuron transfer function, and Fig.~\ref{fig:two-burst}(f) is with a rectifying transfer.

    The fact that input optical signals effect changes in the output optical signal is significant because that output could, in principle, be fed to other neurons; furthermore, the fact that multiple signals can be ``weighted'' by positive and negative values and their sum then influencing the output is an indicator that the MRR neuron can be networked with multiple inputs and outputs. Furthermore, the photodetector-modulator device can be viewed as a wavelength converter with higher gain than all-optical approaches~\cite{Nozaki:18}. Wavelength conversion on a conventional silicon platform has uses outside of neuromorphic photonics.

    The signals themselves were chosen to show very simple signal processing tasks for typical AM radio signals. Linear fan-in is used in RF processing for dimensionality reduction and principal component analysis~\cite{Tait:18ciss}. Followed by a rectifying stage, this system effectively acts as a multi-channel envelope detector (i.e. $(A+B)^2$ and $(A-B)^2$). The expansion of these squares yields a $\pm 2 AB$ term -- a measure of correlation. In radio signal processing, some algorithms for channel estimation are based on calculating correlation to update estimation parameters~\cite{Kutty:16,Tait:18ciss}. Very fast measurements of correlation could accelerate the convergence time and tracking rate of those algorithms.

    % \fxnote{This section was satisfyingly good and convincing. --TFL}

  \subsection{Response to pulses} \label{sec:result-pulses}
    We examine the response of the neuron to two pulsed signals and show that the modulator neuron can be made to exhibit enhancing and saturating nonlinearity as well as inhibition in the pulsed domain. The significance of using pulses is the demonstration of time-resolved processing. When the magnitude of the response is dependent on the precise timing of inputs, the neuron is then capable of implementing tasks that involve temporally-coded signals, for example, temporal pattern detection and polychronization~\cite{izhikevich06}.

\begin{figure}[tb]
      \begin{center}
      \includegraphics[width=0.9\linewidth]{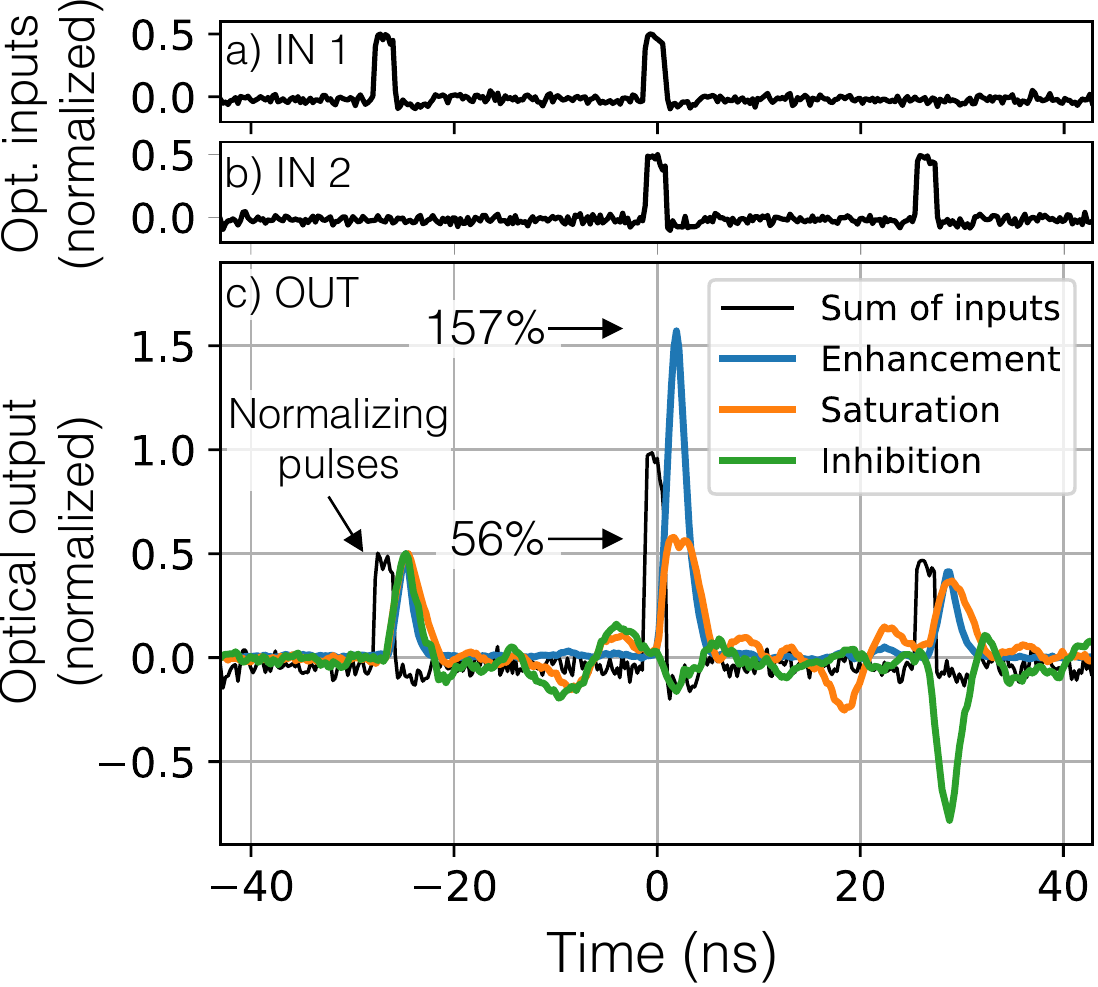}
      \caption[]{Pulse coincidence experimental results. a, b) The input is a 2ns pulse doublet, which is delayed between two input wavelengths to create a pulse coincidence at $t=0$. c) Neuron outputs under different biasing conditions, compared to the linear solution (sum of inputs, black). The first pulse is used as a normalization to 0.5, meaning that the linear sum should peak at 1.0 (i.e. 100\%). Blue: biased for nonlinear enhancement, where coincidence is 157\% of the linear solution. Orange: biased for nonlinear saturation, where coincidence is 56\% of the linear solution. Green: inhibition; fiber connections are switched such that IN 2 is directed to the IN-- photodetector, and the neuron is biased in a linear region. The coincident pulses result in approximately zero output, and the non-coincident pulses cause complementary perturbations (leading, IN 1, positive and lagging, IN 2, negative).}
      \label{fig:pulseCoincidence}
      \end{center}
    \end{figure}

    In this experiment, the input is derived from the PPG producing a pulse doublet with \SI{2.0}{\ns} width and \SI{30}{\ns} (15-bit periods) inter-pulse delay. This signal is modulated onto $\lambda_1$ and $\lambda_2$, which are then delayed separately.
    The inter-channel delay is adjusted to correspond to the inter-pulse delay, such that the middle pulses coincide. At the same time, the leading and lagging pulses show the non-coincident response for comparison. The inputs post-delay are shown in Fig.~\ref{fig:pulseCoincidence}(a, b). Fig.~\ref{fig:pulseCoincidence}(c) compares the enhancing, saturating, and inhibitory cases to the linear sum of inputs (black trace). The traces are normalized to 0.5 at the first non-coincident pulse. This means that the linear sum at coincidence is 1.0.

    Pulse coincidence detection is shown in the blue trace of Fig.~\ref{fig:pulseCoincidence}(c). The coincident pulse peaks at 1.57, meaning that coincidence is over-emphasized. Enhancement is observed when the transfer function corresponds to the ReLU in Fig.~\ref{fig:transferFunctions}(c), biased in the flat region. One pulse is not sufficient to reach the elbow, while two pulses are. We note that the single-pulse suppression is not complete because their bandwidth is near the 3dB bandwidth of the neuron. Coincidence detection is considered to be an essential property of nonlinear units within a pulsed neural network. A positive third-order nonlinearity resulting in pulse compression (shown in Fig.~\ref{fig:1GHzRectify}) is also essential for cascadability in a pulsed neural network because it counteracts pulse width spreading. In the enhancement experiment here, output pulses were not compressed. They have a FWHM of 3.8ns (90\% spread) due to the limited bandwidth of injection modulation. Spreading can be decreased by using longer pulses or faster modulation, such as depletion-mode.

    Pulse saturation is shown in the orange trace in Fig.~\ref{fig:pulseCoincidence}(c). Again, the output is normalized to the leading pulse. In this case, the coincident pulses evoke about the same response as the non-coincident ones. The response corresponds to the inverse ReLU in Fig.~\ref{fig:transferFunctions}(d), biased in the elbow region. The negative deflection around 20ns is a spurious artifact due to carrier relaxation.

    Pulse inhibition is shown in the green trace in Fig.~\ref{fig:pulseCoincidence}(c). Signal IN 2 is switched to the IN-- port of the neuron. The non-coincident pulses can be seen to produce positive (leading pulse) and negative (trailing pulse) effects. When the pulses coincide, the output vanishes. This result indicates more than an inversion of the response from the last experiment, but the simultaneous capability for complementary excitable and inhibitory responses. In a real network, fibers cannot be switched. Instead, wavelength channels can be switched between IN+ and IN-- ports by thermally configurable optical passives~\cite{Tait:16scale}.

  \subsection{Cascadability} \label{sec:cascadability-demo}
    In order for neurons to be cascadable, signals must maintain their integrity from stage to stage indefinitely. The condition of gain cascadability can be expressed as the existence of an operating point where large-signal gain is one and small-signal, differential gain, $g$, is greater than one. The condition of physical cascadability is expressed, for photonic neurons, as the input and output both being optical and at the same wavelength.
    These conditions can be observed using a self-afferent neuron -- also known as an autapse. The autapse is a nonlinear feedback system whose input and output are the same signal by definition.
    As derived in Appendix~\ref{sec:equivalence_bifurcation}, such a system becomes bistable when its differential gain is greater than unity. Therefore, an observation of a transition between mono/bi-stability is equivalent to a demonstration of cascadability. The significance is that bifurcations are qualitative changes readily observable in experiment. Here, we construct an autapse circuit and observe therein this bifurcation, providing evidence that the MRR modulator neuron is indefinitely cascadable.

    \begin{figure}[tb]
      \begin{center}
      \includegraphics[width=0.99\linewidth]{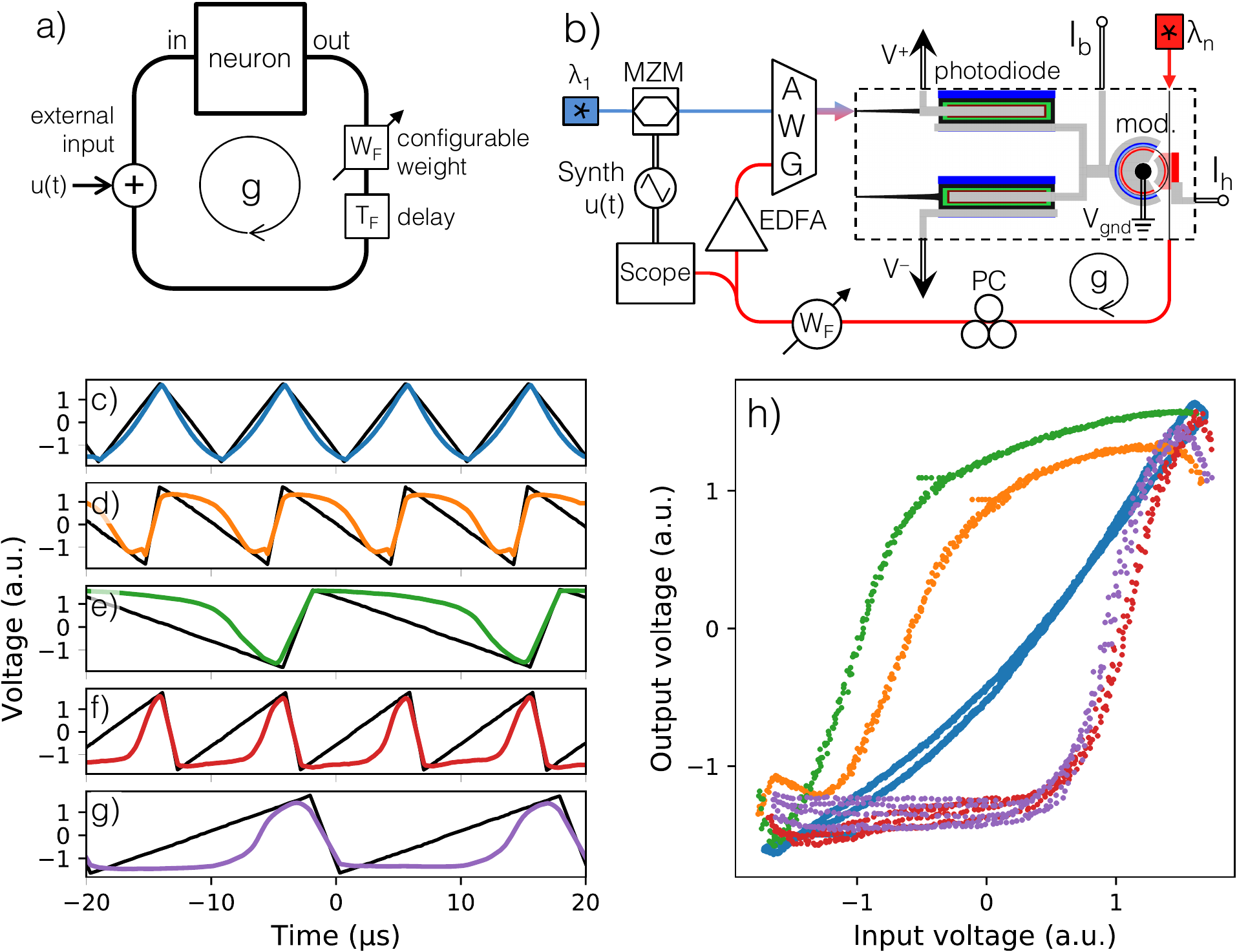}
      \caption[]{The autapse concept and experiment. a) An autapse is a neuron fed back to itself with a unitless round-trip gain of $g$. b) Experimental setup. The output of the neuron on $\lambda_n$ is wavelength multiplexed with a sawtooth modulated external input $u(t)$ on $\lambda_1$. Both are then fed into the positive input photodetector. An oscilloscope records the output and the external input. c) Response to a triangle wave input (black) showing monostability. The feedback attenuator is blocking so that $W_F=0$. d-g) Response with feedback $W_F=1$ and $g > 1$, showing bistability. This is apparent from the difference in falling (d, e) and rising (f, g) edge responses. Changing input from 100kHz~(d, f) to 50kHz~(e, g) effects little appreciable difference. h) Output plotted against input illustrating the opening of a bistable regime. Colors correspond to the output traces in (c--g).}
      \label{fig:autapse}
      \end{center}
    \end{figure}

    The autapse setup is shown in Fig.~\ref{fig:autapse}(b). There is one external input that is modulated by a kHz-timescale sawtooth waveform that is either rising or falling. The output of the neuron at $\lambda_n$ is fed back via fiber and multiplexed with this external signal. Then both are fed into the positive PD. Inline with the feedback pathway is a polarization controller (PC), manually variable feedback attenuator ($W_F$), optical tap to the oscilloscope, and EDFA to counter fiber-to-chip insertion loss.

    Fig.~\ref{fig:autapse}(c) shows the baseline input-output relation for the case where $W_F = 0$. The output (blue) is a regular function of the input, as seen in blue in Fig.~\ref{fig:autapse}(h). Figures~\ref{fig:autapse}(d-g) show the rising and falling cases where $W_F = 1$. The output is different depending on the direction of the input (black). Figure~\ref{fig:autapse}(h) shows each output plotted vs. input, where colors correspond to the left panels. In Figs.~\ref{fig:autapse}(d-g), the biasing state is the same, whereas bias is different in Fig.~\ref{fig:autapse}(c) because the average input optical power is less. These two different qualitative behaviors reveals the presence of a cusp bifurcation between $W_F \in [0,1]$.

    The experimental setup with a fiber in the feedback loop differs substantially from a fully integrated autapse shown in Fig.~\ref{fig:conceptNetwork}. These caveats are shared by~\cite{Tait:17} and are discussed in Sec.~\ref{sec:caveats}.

\section{Discussion}
  \subsection{Claiming cascadability and fan-in} \label{sec:claiming-cascadability}
    Many all-optical and optoelectronic elements exhibit nonlinear or neuron-like input-output transfer functions, but this is not a sufficient condition for said device to be capable of networking with other like devices. Optics faces special challenges in satisfying the critical requirements of cascadability and fan-in, which we will refer to together as network-compatibility. These fundamental challenges were expressed by Keyes and Goodman more than 30 years ago~\cite{Keyes:1985,Goodman:1985}, and they remain challenges today. Above, we have argued that bifurcation in an autapse provides sound empirical evidence of both. Due to their historical difficulty and central importance, we believe that \emph{there should be a high burden of proof placed on proposed photonic neurons to clearly demonstrate cascadability and fan-in.} Here, we will consider how other works have treated these concepts.

    There are several physical properties that pose barriers to cascadability~\cite{Keyes:1985}. In optical devices whose nonlinear mechanism is based on semiconductor carrier-mediated cross-gain modulation (XGM) or Kerr effect, the controlled signal (probe) affects the material properties in the same way as the controller signal (pump). This necessitates weak probes and very small pump-to-probe gains (e.g. the fiber neurons in~\cite{Hill:2002,Rosenbluth:2009}).
    Optical resonators can be used to strengthen these nonlinear effects at the cost of imposing wavelength constraints. In some approaches, output wavelength must be different than the input (e.g. with Kerr~\cite{Xu:07b}, with XGM~\cite{Barbay:11,Nahmias:13,Deng:17}), in which cases, one gate cannot drive another.

    In some recent cases, cascadability is used to refer to a single feedforward connection~\cite{VanVaerenbergh:12,VanVaerenbergh:13,Deng:17,Xiang:17}, but we argue that this approach does not conclusively demonstrate indefinite physical cascadability. The format of the upstream input can be different than that of the downstream output such that the second might not be able to drive the first.
    The autapse provides a stronger demonstration of indefinite cascadability because its output and input signals are the same, and the upstream and downstream neurons are the same by definition. The recurrent circuit is equivalent to a neuron driving an indefinite feedforward chain of other identical neurons.

    Some have approached cascadability by avoiding all-optoelectronic signal pathways and instead using an O/E/O chain consisting of a photodetector connected to a laser~\cite{Nahmias:16,Shainline:18II,Shainline:18IV} or modulator~\cite{Tait:17,Nozaki:18}. Wavelength constraints and phase sensitivity vanish because this information is lost in the electronic domain. The E/O conversion step can offer strong nonlinearity, as employed here, and the electronic domain itself offers efficient mechanisms for nonlinearity and amplification. In~\cite{Shainline:17}, an O/E/O neuron based on cryogenic silicon LEDs, superconducting detectors, and superconducting amplifiers~\cite{McCaughan:14} was proposed. Its physical cascadability was demonstrated by the E/O/E LED-detector link shown in~\cite{Buckley:17jrnl}, and its gain cascadability has been addressed in more recent theoretical works~\cite{Shainline:18II,Shainline:18IV}. The downside of O/E/O is a vulnerability to electrical parasitics; however, these parasitics can remain small regardless of network scale because O/E/O occurs entirely within a neuron, not between neurons.

    % \fxwarning{this is intriguing but breaks up the flow}
    % Weighted fan-in needs some way to put values originating from different places into one summed value. There are many disparate instances of fan-in in electronics. In CPUs, a register accumulates values from different memory addresses. In neuromorphic electronics, registers accumulate values arriving via a distributed packet network. In CMOS logic gates, inputs arrive at different transistors that are arranged in series or parallel. In analog vector-matrix multipliers, it is charge starting on multiple switched capacitors equilibrating over the junction between them, or it could be current starting in two wires, leaving from a third.

    When light combines, it interferes, posing a fundamental challenge to fan-in~\cite{Goodman:1985}. Optical fan-in results in either phase-dependence, when coherent, or $N$-fold loss, when incoherent (e.g. 3dB at $N=2$). In some all-optical devices where the in/out wavelengths can be the same (cascadable), these wavelengths also must be the same, meaning they cannot have more than one input~\cite{VanVaerenbergh:12,VanVaerenbergh:13}. Fan-in with coherent signals can be achieved by exerting complete control of the optical phase in the interconnect~\cite{Shen:17}, but then signal-dependent phase changes in a neuron profoundly affect the behavior of the subsequent interconnect, precluding any cascadability. In~\cite{Shen:17}, neuron calculations were implemented at low-speed in a CPU; a neuron based on saturable absorption was contemplated, but it was not discussed how this element would regenerate a consistent optical phase.

    Fan-in has also been achieved using inputs that are coherent, but mutually incoherent, such as different spatial modes~\cite{McCormick:93,Fok:11a}, different polarizations, or different wavelengths~\cite{Khan:2010,Chang:14implement,Tait:15}. These signals do not interfere, and, since they are individually coherent, can be multiplexed and routed/weighted independently by tunable resonators~\cite{Tait:16multi,Xu:11}. Total power is sensed by a photodetector (O/E), making this fan-in approach compatible with the O/E/O approach to cascadability. Multi-wavelength weighted addition was combined with O/E/O laser neurons in~\cite{Nahmias:16,Peng:18}, wherein cascadability was also considered but not directly demonstrated. A downside of relying on multiple wavelengths is the need for a different laser source for each channel. The size of a single all-to-all subnetwork is capped by the available spectrum and the ability to distinguish adjacent channels, found in~\cite{Tait:18twopole} to be less than 950 if using the resonators of~\cite{Soltani:10}; however, multiple of these subnetworks could be interfaced on a single chip~\cite{Tait:14}.

  \subsection{Non-spiking photonic neurons} \label{sec:departing-from-spiking}
    The great majority of work on photonic neurons has focused on lasers that implement spiking models similar to biological neurons~\cite{Nahmias:13,coomans2011solitary,VanVaerenbergh:13,Brunstein:2012,Selmi:2014,Romeira:16,Nahmias:16,Deng:17,Peng:18}, reviewed in~\cite{Prucnal:16advances}. To claim a non-spiking modulator as a photonic neuron represents a departure from the origins of neuromorphic photonics. Modulator-class photonic neurons are easier to fabricate and dissipate less power on-chip. On the other hand, the processing repertoire of spiking neurons is theoretically greater than continuous-valued neurons.

    The question is whether a move to continuous-valued photonic neurons would forfeit too much processing richness.
    Spiking networks can perform any continuous-valued task by representing values as mean firing rate (MFR). These are converted to analog signals by low-pass filtering. A mathematical derivation of the MFR transform is found in~\cite{Tang:17}. In many applications, a valid alternative is to use physically analog signals to represent the MFR directly. There is no debate over whether continuous-valued neurons have numerous engineering applications. For example, in traditional artificial neuromorphic computers, the spike rate is represented by analog voltage~\cite{rosenblatt58,Neumann:56}. Modern machine learning almost universally makes the MFR transform, representing MFR as floating-point numbers. Another example is the NEF compiling process~\cite{stewart2014large} that only looks at MFR-to-MFR transfer functions. By replacing the MFR transfer functions with analog optical-to-optical transfer functions, we found that the whole NEF framework is still applicable to networks of modulator neurons~\cite{Tait:17}.

    There are at least four situations where spiking artificial neurons have essential motivation; without these motivations, continuous-valued neurons are just as applicable. These motives include 1) studying brain function~\cite{Friedmann:13,Diamond:16}, 2) a need for increased robustness to amplitude noise, and 3) the use or study of temporal coding. Temporal coding has been proposed as one of the keys to energy efficiency in biological and perhaps artificial neural networks~\cite{Sarpeshkar:98}; however, this proposition is debated and the subject of active theoretical research (generally originating from~\cite{Maass:97}). Finally, a need for spiking can stem from 4) hardware constraints. The modern wave of electronic neuromorphic hardware is based exclusively on spiking neural models because metal wires perform poorly when implementing dense, analog interconnects. Digital packet routing is used to multiplex a small number of physical interconnects, creating a large number of virtual interconnects~\cite{Furber:14,Akopyan:15}. In these cases, packets are treated as spikes.

    There is no fundamental reason why photonics must have one of these motivations, so we believe that continuous-valued modulator neurons do not sacrifice too much computational richness. We postulate that, due to their complementary nature, laser and modulator neurons will come to address complementary application domains. It is even possible that they could exist on the same chip to handle different types of tasks within a single neural network.

  \subsection{Caveats of the fiber-based autapse} \label{sec:caveats}
    Both here and in~\cite{Tait:17}, the feedback loop used for an autapse partially consists of fiber, resulting in substantial gain, frequency, and robustness discrepancies between the experiment and a fully integrated autapse. The high insertion loss of fiber-to-chip couplers necessitate an EDFA in the feedback loop, weakening the claim that we have directly demonstrated gain cascadability. Direct characterization showed a gain of 1.36, accounting for fiber-to-chip insertion loss (Sec.~\ref{sec:characterization}); device-level calculations indicate that, without fiber-to-chip insertion loss, $g$ can readily exceed unity.

    The kHz frequencies of the input sawtooths in Fig.~\ref{fig:autapse}(c-g) are chosen to isolate mono/bi-stable equilibrium effects from dynamical effects. If the frequency were too high, then the long fiber feedback delay would lead to obfuscating time-delayed dynamics~\cite{Romeira:14}. On the other hand, a modulation bandwidth that is too slow will also obscure equilibrium dynamics governed by carrier injection modulation. The thermal dissipation time constant is near 1ms, and the EDFA gain population timeconstant is 9ms. The temperature and population states effectively act as integrators around 1kHz and 100Hz, respectively. The resulting phase lag in the output closely resembles bistability. To eliminate the possibility that we are actually observing phase lag dynamics, two different frequencies (50kHz and 100kHz) are compared in Fig.~\ref{fig:autapse}(d vs. e and f vs. g). The fact that they produce very similar responses provides evidence that we are observing equilibrium effects, as opposed to dynamic effects that are frequency-dependent.

    The fiber setup has poor environmental robustness. The couplers are polarization-sensitive, and the feedback fiber can drift. This means the polarization controller must be tuned relatively often to maintain a consistent round-trip gain. The average optical power coming out of the modulator is non-zero, thereby affecting the steady-state bias on the modulator. To further complicate that effect, the average power is weighted by the same amount as the varying signal (see Eq.(5) of Ref.~\cite{Tait:16scale}). To counter this effect, the prior work~\cite{Tait:17} used an in-line capacitor as a DC block. That is not possible here because the key PD-modulator junction is fixed on-chip, so we perform this correction manually for the two feedback weight values. In a real network, it would be relatively straightforward to calculate the bias deflection due to weighted average power and then to correct the applied bias current to result in the desired net bias. We leave the automated version of that control algorithm to future work.

    The fiber-based autapse has intriguing resemblances to photonic reservoir computers based on time-delayed feedback~\cite{Larger:12,Hermans:15,Ortin:15,Soriano:15,Romeira:16} although they represent distinct approaches to computing. In fiber reservoirs, the feedback fiber is used to induce complex dynamics, unlike autapse demonstrations in which we attempt to minimize these dynamics in order to show isomorphism to a simple model. The reservoir concept complements neural network approaches, as they are typically understood, precisely because it does not rely on maintaining a correspondence between hardware and a theoretical model. This means reservoirs can employ continuous substrates that are easy to fabricate but difficult or impossible to make adhere to a model, such as optical phase around a fiber or wave amplitude in a bucket of water~\cite{Fernando:03}. A network of neurons can also, of course, act as a reservoir by remaining ignorant to its theoretical model~\cite{Maass:02}. Reservoir computers rely on instance-specific training to isolate dynamics desirable for processing -- there is no guarantee that a particular hardware instance will exhibit the desired dynamics at all. In contrast, any artificial system isomorphic to an abstract neural network is guaranteed to be capable of all dynamics possible in that model.

  \subsection{Further work}
    The results of this paper pose questions for further research. It is apparent how the modulator neuron introduced in this work could be co-integrated with the microring weight of Ref.~\cite{Tait:17}, together shown in Fig.~\ref{fig:conceptNetwork}. An important demonstration for future work will be a fully-integrated autapse. An integrated autapse would not suffer from fiber-induced latency or fiber-to-chip coupling loss. As hypothesized in~\cite{Tait:18thesis}, an integrated autapse could provide a means to accurately and experimentally quantify the energy consumption of neuromorphic photonic systems.

    Further work should look to reproduce these results at higher bandwidth. There were two bandwidth limiters: 1) the use of carrier injection modulation and 2) no isolation from the capacitance of the $I_b$ connection. The first limitation could be addressed by optimizing the MRR modulator for stronger depletion modulation. The second could be addressed by an on-chip resistor and series inductor. The capacitance of the modulator junction is affected by the metal between it and the dominant impedance of the biasing source. In this work, that included the junction itself, its breakout trace to the pad, the probe contacting the pad, and the wire leading to the Keithley. An on-chip resistor could constrain this capacitive region to just the junction itself.

    Further work could also explore computational behavior by combining the experimentally validated transfer functions of the MRR modulator neuron with existing neural algorithms and compilers~\cite{stewart2014large}. These compilers take a neuron transfer function and a high-level task specification, then returning a weight matrix. By mapping these weights back to hardware, the physical neural network will perform the task. This is analogous to a program compiler that takes a computer architecture and a program written in a high-level language, then returning machine code. Prior work~\cite{Tait:17} demonstrated this strategy in photonics using the Neural Engineering Framework (NEF)~\cite{Eliasmith:04}. Likewise, the validated transfer functions could be injected into algorithms for machine learning~\cite{George:18}.

  % \fxwarning{where do we talk about control and learning}

\section{Conclusion}
  We fabricated a silicon photonic circuit consisting of two photodetectors electrically connected to a microring modulator. We then directly demonstrated its ability to act as a network-compatible photonic neuron, namely the properties of optical-to-optical nonlinearity, fan-in, and indefinite cascadability. These properties have never been demonstrated together in an integrated optoelectronic device.
  % This demonstration was limited in several respects yet shows directions for further research. Since this modulator was based on carrier injection, its feedforward behavior was much slower than it could be if using carrier depletion. Since weights were implemented off-chip, its feedback behavior was much slower than if using co-integrated weights. Only one neuron was examined, its potential to act within a processing network of similar neurons considered in detail with demonstration posed to future work.

  Combined with microring weight banks shown in~\cite{Tait:17,Tait:18fb}, the modulator neuron constitutes the final component needed to implement broadcast-and-weight~\cite{Tait:14} neural networks: to date, the only architecture to propose a way to unite well-defined photonic neurons and well-defined neural interconnects upon a mainstream silicon photonic platform. Compatibility with silicon photonics provides crucial aspects of feasibility, scalability, and economies of scale to modern day photonic neural networks. Due to their unprecedented potential speed, neuromorphic photonic systems can come to bear on society's growing demand for datacenter machine learning and could, additionally, open up unexplored regimes of ultrafast machine intelligence.

\section{Appendix: equivalence of cascadability and bifurcation} \label{sec:equivalence_bifurcation}
  Proving that particular piece of silicon exhibits a mathematical isomorphism to a desired model is not trivial, even for a system as simple as a single neuron. This proof is important because, if and only if a device is isomorphic to a model, can that model be used to predict behaviors of larger networks of those devices. Our approach to a rigorous photonic neural network demonstration is to derive qualitative predictions of a characteristic model, and then to observe them in experiment. This observability is crucial for connecting abstraction to reality. In Sec.~\ref{sec:cascadability-demo}, we showed a mono/bi-stable bifurcation in an autapse in attempts to demonstrate cascadability. Here, we derive the equivalence between the cascadability condition and a bifurcation threshold, only the latter of which is an observable phenomenon.

\begin{figure}[tb]
    \begin{center}
    \includegraphics[width=0.6\linewidth]{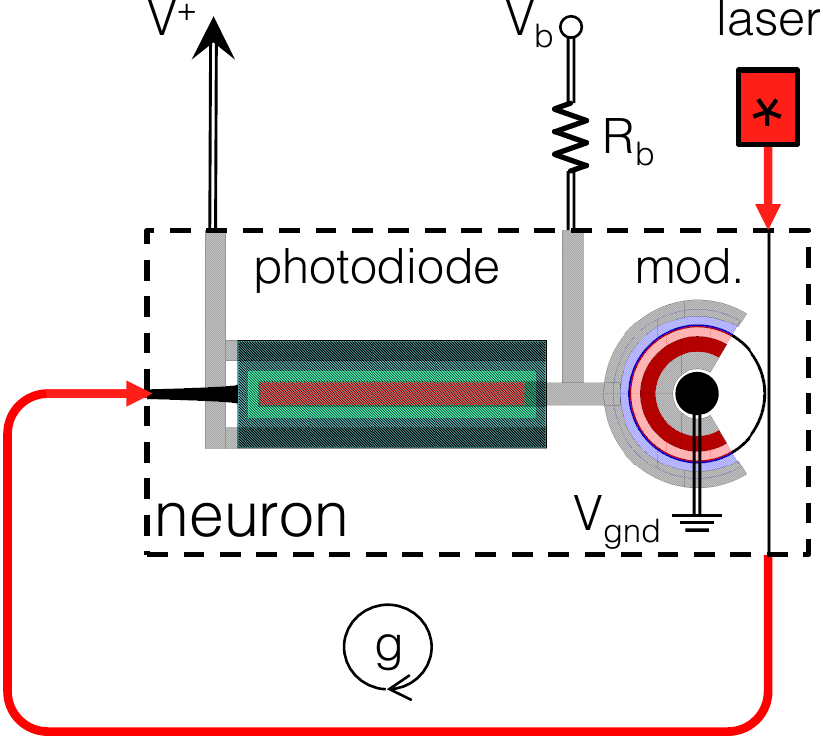}
    \caption[]{A photonic autapse implemented by a microring modulator neuron. The modulator junction is reverse biased by a voltage source with impedance $R_b$.}
    \label{fig:modulator_autapse}
    \end{center}
  \end{figure}

  The modulator neuron autapse is shown in Fig.~\ref{fig:modulator_autapse}. We assume that a voltage-mode depletion modulation effect is used instead of current-biased injection modulation as above because it will eventually be the preferred approach.
  The cascadability condition is met when the neuron's differential optical-to-optical gain, $g$, exceeds unity.
  \subsection{Theory}
    $g$ can be derived from device properties. It is defined
    \beq \label{eq:cascadability-condition}
      g &=& \frac{dP_{out}}{dP_{in}} \\
      &=& \frac{dP_{out}}{dT_{mod}} \frac{dT_{mod}}{dV} \frac{dV}{dP_{in}}
    \eeq
    where $T_{mod}$ is modulator transmission, and $V$ is the junction voltage.

    The power leaving the modulator is
    \beq
      P_{out} &=& P_{pump} T_{mod}(V) \\
      \frac{dP_{out}}{dT_{mod}} &=& P_{pump}
    \eeq
    where $P_{pump}$ is the optical power entering the modulator, and $V$ is the junction voltage. We assume that the modulator is biased at its maximum slope point for optimal modulation slope efficiency. When there are no inputs, $V = V_b$, where $V_b$ is the bias voltage. The optimal bias voltage is $V_b^*$.
    \beq
      V_b^* &=& \argmax_{V} \frac{dT_{mod}(V)}{dV} \\
      % T_{mod}(V) &\approx& T_{mod}(V^*) + \left.\frac{dT_{mod}(V^*)}{dV}\right|_{V^*} (V - V^*) \\
      \left.\frac{dT_{mod}}{dV}\right|_{V_b = V_b^*} &=& \frac{\pi}{2 V_{\pi}}
    \eeq
    where $V_{\pi}$ is called the $\pi$-voltage, used for notational convenience. $V_{\pi}$ is inversely proportional to $dT_{mod}/dV$. It means the voltage to induce a $\pi$ shift in the Mach-Zehnder modulator with equivalent modulation slope efficiency, not the voltage to induce a $\pi$ shift around the MRR.

    The detector and receiver component is
    \beq
      V &=& V_b + R_{pd} R_b P_{in} \\
      \frac{dV}{dP_{in}} &=& R_{pd} R_b
      % I_{in} &=& R_{pd} P_{in}
    \eeq
    where $R_{pd}$ is detector responsivity and $R_b$ is the biasing impedance. Combining the above terms,
    \beq
      g &=& P_{pump} \cdot \frac{\pi}{2 V_{\pi}} \cdot R_{pd} R_b \\
      \left.P_{pump}\right|_{g=1} &=& \frac{2 V_\pi}{\pi R_{pd} R_b} \label{eq:pump-power}
    \eeq
    This pump power is the minimum needed for cascadability.

  \subsection{Observation}
    The physical autapse circuit can be modeled as a dynamical system whose state variable is the modulator voltage. A bifurcation is a qualitative change that occurs when the number or stability of fixed points changes. In the Supplementary Material of~\cite{Tait:17}, the fixed points were derived with a simplified model. Here, we examine fixed point stability in a physical model and find the bifurcation point to be identical to the expression in Eq.~\ref{eq:pump-power}.

    The modulator junction has these dynamics
    \beq
      \frac{dV}{dt} &=& - \frac{V - V_b}{R_b C_{mod}} + \frac{i_{pd}}{C_{mod}}
    \eeq
    where $C_{mod}$ is the capacitance of the modulator junction, and $i_{pd}$ is the photocurrent. Photocurrent is induced by either the external signal, $P_{ext}$, or the self-feedback signal, $P_{in}$. Due to the feedback, $P_{in} = P_{out}$.
    \beq
      i_{pd} &=& R_{pd} \left[P_{out}\left(V\right) + P_{ext}\right]
    \eeq

    Zeroing the external input, this results in
    \beq
      \frac{dV}{dt} &=& - \frac{V - V_b}{R_b C_{mod}} + \frac{R_{pd}}{C_{mod}} P_{pump} T_{mod}\left(V\right) \label{eq:full_physical}
    \eeq
    We do not seek the solutions for the steady-state values because they have an inelegant dependence on the parameters. Instead, we assess the stability using the Jacobian linearization. In general, dynamical system are described by $\dot{x} = \mathbf{F}(x)$, where $\mathbf{F}$ is a matrix of functions. They can be linearized around fixed points, $x_{fp}$, as
    \beq
      \lim_{\Delta x \to 0} \Delta \dot{x} &=& \mathbf{J} \Delta x \\
      \mbox{where } \ \mathbf{J} &\equiv& \left.\frac{\partial \mathbf{F}}{\partial x}\right|_{x_{fp}}
    \eeq
    where $\Delta x = x - x_{fp}$ and $\mathbf{J}$ is called the Jacobian matrix. In our one-dimensional case, the Jacobian is just a scalar, $J$. When it is positive $\Delta x$ will grow exponentially, meaning that the fixed point is unstable. A bifurcation occurs when stability changes, meaning when $J = 0$. Of course, the exponential growth saturates at one of two other fixed points. In other words, in a 1-dimensional system, this bifurcation is a transition from monostability to bistability.

    Using the physical description of the autapse dynamical system from Eq.~\eqref{eq:full_physical},
    \beq
      J &=& - \frac{1}{R_b C_{mod}} + \frac{R_{pd}}{C_{mod}} P_{pump} \left.\frac{dT}{dV}\right|_{V_{b}^*}
    \eeq
    which crosses zero when
    \beq
      \left.P_{pump}\right|_{J=0} &=& \frac{2 V_\pi}{\pi R_{pd} R_b} \label{eq:destabilization}
    \eeq
    Thus, the expression for pump power where the autapse loses monostability corresponds exactly with that where the cascadability condition is met in Eq.~\eqref{eq:pump-power}.
    % This is significant because Eq.~\eqref{eq:pump-power} was calculated from theory, while Eq.~\eqref{eq:destabilization} is observable in experiment as a transition from monostability to bistability.

\section*{Funding}
  National Science Foundation (NSF) (ECCS 1247298, DGE 1148900)

\section*{Acknowledgment}
  Devices were fabricated at the IME A$^*$STAR foundry in Singapore. Fabrication support was provided via the Natural Sciences and Engineering Research Council of Canada (NSERC) Silicon Electronic-Photonic Integrated Circuits (SiEPIC) Program and the Canadian Microelectronics Corporation (CMC).

\bibliographystyle{apsrev4-1}
%% Step 1: Compile from a Master BibTex
% \bibliography{Master_Biblio}
%% Step 2: In terminal, run ./bibScript
%% Step 3: Compile from this project BibTex
\bibliography{mrrMod}

\end{document}